\documentclass[aps,pra,twocolumn,groupedaddress,superscriptaddress,nofootinbib]{revtex4-1}
\usepackage{graphicx}

\usepackage{amsmath}
\usepackage{amsfonts}
\usepackage{amssymb}
\usepackage{dcolumn}
\usepackage[normalem]{ulem}
\usepackage{enumerate}
\usepackage{enumitem}
\usepackage{bbold}
\usepackage{cancel}
\usepackage{mathtools}
\usepackage{siunitx}
\usepackage{braket}
\usepackage{amsmath}
\usepackage{mdframed}
\usepackage{soul, color}

\usepackage{algorithm}
\usepackage[noend]{algpseudocode}

\usepackage{lipsum}

\usepackage[usenames,dvipsnames]{xcolor}
\usepackage[colorlinks,bookmarks=false,citecolor=NavyBlue,linkcolor=OliveGreen,
urlcolor=blue]{hyperref}

\renewcommand{\vec}[1]{\boldsymbol{#1}}

\newcommand{\sandwich}[2]{\left<\left.#1\, \right| #2\, \right>}
\newcommand{\II}{\mathrm{i}}
\newcommand{\D}{\mathrm{d}}
\newcommand{\norm}[1]{\left\lVert#1\right\rVert}
\newcommand{\proj}[2]{\ket{#1}\hspace{-.15cm}\bra{#2}}

\definecolor{gruen}{HTML}{006800}
\definecolor{darkorange}{RGB}{255,91,45}

\begin{document}
\title{From classical to quantum walks with stochastic resetting on networks}
\author{Sascha Wald}
\email{swald@pks.mpg.de}
\thanks{Both authors contributed equally to this work.}
\affiliation{Max-Planck-Institut f\"ur Physik Komplexer Systeme,
N\"othnitzer Stra{\ss}e 38, D-01187, Dresden, Germany\looseness=-1
}

\author{Lucas B\"ottcher}
\email{lucasb@ethz.ch}
\affiliation{Computational Medicine, UCLA, 90024, Los Angeles, United States}
\affiliation{Institute for Theoretical Physics, ETH Zurich, 8093, Zurich, Switzerland}
\affiliation{Center of Economic Research, ETH Zurich, 8092, Zurich, Switzerland}
\date{\today}
%
%
%
\begin{abstract}
Random walks are fundamental models of stochastic processes with applications in various fields including physics, biology, and computer science. We study 
classical and quantum random walks under the influence of stochastic resetting on 
arbitrary networks. Based on the mathematical formalism of quantum stochastic 
walks, we provide a framework of classical and quantum 
walks whose evolution is determined by graph Laplacians. We
study the influence of quantum effects on the stationary and long-time 
average probability distribution by interpolating between the classical and 
quantum regime. We compare our analytical results on stationary and long-time 
average probability distributions with numerical simulations on different 
networks, revealing differences in the way resets affect the sampling properties of classical and quantum walks.
\end{abstract}
\maketitle

\section{Introduction}
\noindent
Karl Pearson coined the term ``\textit{random walk}'' in a short commentary 
article in 1905~\cite{pearson1905problem}. In the same 
year Albert Einstein described the random movements of particles suspended
in a fluid in terms of Brownian motion~\cite{einstein1905molekularkinetischen,kac1947random}, 
illustrating the potential of stochastic descriptions 
to improve our understanding of physical processes.

 Early versions of random walks were also applied to problems in 
 probability theory~\cite{epstein2012theory} or 
 in materials science~\cite{flory1969statistical,rubinstein2003polymer,gujrati2010modeling}.
 Nowadays, applications of random walks are quite versatile and are used to describe stock-price 
 fluctuations~\cite{mandelbrot1999multifractal,malkiel1999random},
 foraging animals~\cite{maarell2002foraging,bartumeus2005animal},
 efficient search algorithms~\cite{gkantsidis2004random} such as the famous PageRank~\cite{page1999pagerank} 
 or even opinion formation~\cite{degroot1974reaching,bottcher2020competing,bottcher2020great}. 
We could continue this list but instead refer the interested reader to 
Refs.~\cite{ben2000diffusion,noh2004random,masuda2017random} for further information.

Naturally, the success of \textit{classical random walks} (CRWs) 
\begin{figure*}[t!]
\includegraphics[width=\textwidth]{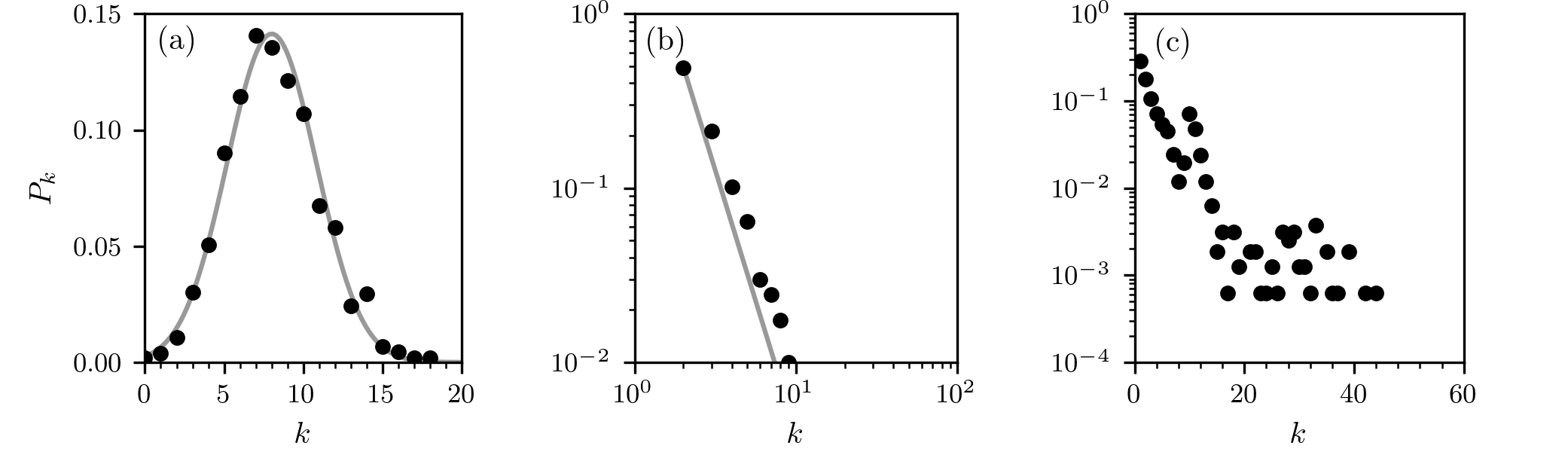}
\caption{\textbf{Degree distributions of different networks.} We show the degree distributions of the different networks that we use throughout this work.
Black disks represent the degree distribution of a particular network realization and the gray solid line is the corresponding analytic degree distribution.
(a) Erd\H{o}s-R\'enyi network with a Gaussian degree distribution and $N=1600$ nodes.
(b) Barab\'asi-Albert network with a power-law degree distribution. The corresponding exponent is $-3$~\cite{albert2002statistical}.
\textcolor{black}{Each new node is connected to $m=2$ existing nodes (i.e., the degree of each node is at least 2) and 
the depicted realization has $N=1600$ nodes.}
(c) Peer-to-peer network (p2p-Gnutella08)~\cite{snapnets} with $N=1600$ nodes and mean degree $\bar{k} \approx 2.5$.
}
\label{fig:networks}
\end{figure*}
has stimulated intensive research on the 
implementation and possible applications of quantum versions of random walks.
Such \textit{quantum walks} (QWs) have been realized in various experimental setup, cf.~\cite{Ryan05,karski2009quantum,Schmi09,Zaeh10,Bouw99,Do05,Schr10,Pere08}.
Although single particle QWs are usually closely related to classical wave phenomena~\cite{Knight03}, quantum extensions~\cite{watrous2001quantum} 
led to important advances such as
the development of hybrid  classical-quantum versions of the 
PageRank~\cite{Buri12,Lecca19}, new 
classes of quantum 
algorithms~\cite{shenvi2003quantum,childs2004spatial,tulsi2008faster,
magniez2011search}, and important insights into the feasibility of quantum 
computations in terms of QWs~\cite{childs2009universal}.\footnote{Note that different formulations of QWs have been proposed, including coined 
QWs~\cite{aharonov1993quantum,aharonov2001quantum}, or the 
formalism developed by Szegedy~\cite{szegedy2004quantum}. 
Connections between coined 
and Szegedy QWs have been established through staggered 
QWs~\cite{portugal2016staggered,portugal2016staggered2}. For an overview of QWs 
and search algorithms, see Ref.~\cite{portugal2013quantum} and for yet other 
formulations of QWs see 
Refs.~\cite{leung2010coined,kollar2012asymptotic,chandrashekar2014quantum,
elster2015quantum,portugal2016staggered,ghosal2018quantum}.}

To use general walks in optimization problems it may be advantageous to employ 
algorithms with \textit{stochastic resetting}~\cite{evans2011diffusion,evans2020stochastic}, since optimization strategies without 
reset may end up in regions far away from the actual 
solutions~\cite{luby1993optimal,montero2017continuous}.
Stochastic resetting is relevant not only in the context of search strategies, but also useful to sample certain rare events in simulations~\cite{villen1991restart}.
Furthermore, stochastic resetting has been theoretically analyzed in a tilted Bose--Hubbard system~\cite{mukherjee2018quantum} for which 
results have been shown to smoothly interpolate between results obtained from the diagonal ensemble (vanishing reset rate) and the quantum Zeno effect (large reset rate). In a very recent work~\cite{riascos2020random}, classical random walks with stochastic resetting were studied on networks. For Caley trees, it was found that a classical random walker with an optimized stochastic resetting protocol performs similarly well as optimal search strategies in finding a target node at a certain distance. In other networks, the performance of such optimized random-walk searches depends on the centrality of the node that is chosen as reset state~\cite{riascos2020random}.

Motivated by the success of random-walk-based search 
strategies with stochastic resetting~\cite{ham2013generalized,valdeolivas2019random}, we study stochastic quantum walks with resetting on networks. To do so, we introduce a
framework for classical, quantum, and hybrid random walks with stochastic resetting on networks. Our approach enables us to (i) interpolate between the classical and quantum regimes of such walks and (ii) analyze the influence of classical processes that perturb QWs. 
We explicitly show how a hybrid classical-quantum walk can be defined as an open quantum system and how the quantum jump operators and dissipation rates are connected to the CRW Hamiltonian. 
From this formulation, we derive the node occupation statistics in the stationary state of classical and quantum walks with stochastic resetting.
We complement these efforts with a detailed numerical analysis for which we unravel the quantum master equation as stochastic Schr\"odinger equation 
and derive the adequate quantum jump probabilities. All codes that we used in this work are made publicly available~\cite{git_walk}.

The paper is organized as follows. In Sec.~\ref{sec:networks}, we briefly review 
some key concepts from the study of complex networks
that will be used in our further analysis. In Sec.~\ref{sec:qsws}, we 
provide precise definitions of the CRWs and the QWs we consider and 
understand these walks as cornerstones for a linear interpolation scheme that 
allows us to add ``classicality'' to QWs and vice versa ``quantumness'' to 
CRWs. 
Our general formulation of classical, quantum, and hybrid classical-quantum walks on networks 
utilizes the theoretical framework of quantum stochastic walks~\cite{Whit10}, 
which we adapt to account for stochastic resetting.
We compare our analytical results with corresponding simulations on different 
networks in Secs.~\ref{sec:numerical_recp} and \ref{sec:numerics}. In 
Sec.~\ref{sec:discussion}, we discuss our results and conclude.
%
%
%

\section{Network science concepts}
\label{sec:networks}
\noindent
A graph $G$ (i.e.,~a network) is an ordered pair of two sets 
$G=(V,E)$ where $V$
is the set of {\it nodes} and $E \subseteq V \times V$ is the set of {\it 
edges}, 
respectively~\cite{newman2018networks}.
We denote the number of nodes by $N$ (i.e., $ |V| = N $). 
Throughout this paper, 
we consider undirected networks, meaning that all edges 
are bidirectional, with unweighted edges.
In order to describe dynamics on such graphs, we introduce a Hilbert space 
structure in the standard way
by assigning to each node $i$ a basis vector $\ket{i}$. These basis 
vectors are chosen to be orthonormal, i.e.~$\sandwich{i}{j} = \delta_{ij}$.
The {\it adjacency matrix}
$A$ of a graph $G$ describes the connections of the graph.
For undirected networks with unweighted edges, each matrix element 
$A_{i j}\in \{0,1\}$ is
\begin{align}
A_{i j} = \begin{dcases}
           1, &\text{if}\ (i,j)\in E \, , \\
           0, &\text{otherwise}\,.
          \end{dcases}
\end{align}
The adjacency matrix is thus a symmetric binary matrix and can be written in 
Dirac notation as
\begin{equation}
A=\sum_{i, j \in V} A_{i j} \ket{i} \bra{j} \ .
\end{equation}
The {\it degree} $k_i$ of node $i$ is defined as the sum over the respective row 
of the adjacency matrix (i.e., $k_i=\sum_{j=1}^N A_{ij}$) and counts the number 
of connections leading to (respectively away from) node $i$. 

A characteristic quantity for graphs is the {\it degree distribution} that indicates
the frequency of nodes with a certain degree. It is defined as
$P_k=n_k/N$, where $n_k$ is the number of nodes of degree $k$ in $G$. The {\it 
degree matrix} $D$ of $G$ is
\begin{equation} \label{eq:D}
D=\sum_{i=1}^N k_i \ket{i} \bra{i} \, .
\end{equation}
$D$ is a diagonal matrix whose elements correspond to the 
degree of the respective node.
The evolution of CRWs and QWs on a graph $G(V,E)$ can be described by the 
{\it graph Laplacian}~\cite{faccin2013degree}
\begin{equation}\label{eq:L}
\mathbb{L}=D-A\, , \quad  \mathbb{L}_{ij}=\begin{dcases}
                  k_i,\, &\text{if}\quad i=j\\
                  -1,\, &\text{if}\quad (i,j)\in E\\
                  0, \, &\text{otherwise}
                 \end{dcases}
\, .
\end{equation}

These theoretical tools suffice for the study of the types of random walks that
we envision. We shall now specify the set of graphs that we use in the present work.
While our analytic results can be applied to general graphs, we will focus on 
the following three different networks
for explicit numerical verification of our results.
\begin{itemize}
 \item  \textbf{Erd\H{o}s-R\'enyi}:
        Two nodes are connected with probability $p$, which is independent of 
all other connections. Erd\H{o}s-R\'enyi graphs have a 
        binomial degree distribution~\cite{newman2018networks}.
        
 \item  \textbf{Barab\'asi-Albert}: \textcolor{black}{A new node will be attached to $m \leq m_0$ 
existing nodes and the attachment probability is proportional to the number of 
edges of the
existing nodes. Here $m_0$ is the initial number of nodes. This preferential-attachment process leads to a scale-free 
network with an algebraic degree distribution~\cite{albert2002statistical}.}
        
 \item  \textbf{Peer-to-peer (p2p-Gnutella08)}: Nodes in this empirical network 
correspond to computers in a file-sharing network~\cite{snapnets}.
\end{itemize}
In Fig.~\ref{fig:networks} we show the degree distributions and exemplary 
realizations of these networks.
Our choice of the outlined networks is motivated by their different connectivity patterns, which enable us to study how such differences affect the properties 
of CRWs and QWs.
\begin{figure*}
\centering
\includegraphics[width=0.75\textwidth]{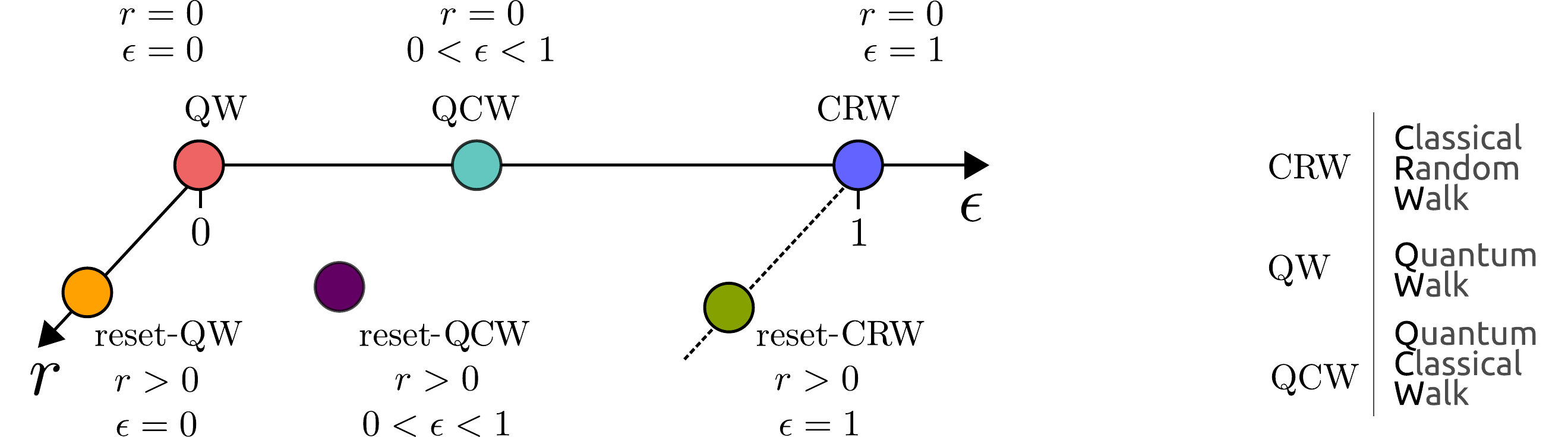}
\caption{\textbf{Overview of different classical and quantum reset walks.} Different types of random walks as function of the classicality parameter 
$\epsilon$ (see Eq.~(\ref{eq:qsw})) and reset rate $r$ (see Eq.~(\ref{eq:reset})). The classicality parameter $0\leq\epsilon\leq1$ allows us to 
smoothly interpolate between a QW ($\epsilon = 0$) and a CRW ($\epsilon =1$). In the time interval $[t,t+\D t]$, a reset to the initial state occurs with probability $r \D t$.}
\label{fig:walks_overview}
\end{figure*}
%
%
%
\section{Quantum Stochastic Walks on networks}
\label{sec:qsws}
\noindent
In this section, we shall specify the types of walks on graphs that we study in 
the present work. Our choice is not unique and we refer the reader to 
Refs.~\cite{boettcherhermannbook,portugal2013quantum}
for an overview of different realizations of CRWs and 
QWs. The walks that we study in this work are 
schematically summarized in Fig.~\ref{fig:walks_overview}.

We start by introducing CRWs (Sec.~\ref{sec:crw}) in terms of their probability 
distribution and 
QWs (Sec.~\ref{sec:qrws}) in terms of the corresponding wave function. 
CRWs and (continuous) QWs are fundamentally different, as the latter obey unitary dynamics as long as 
the walker is not measured, whereas the former do not.
It has been argued that the characteristics of both types of dynamics can be 
incorporated in the dissipative dynamics of the density matrix of a quantum 
system, resulting in quantum 
stochastic walks~\cite{Whit10}. General quantum stochastic
processes can be described by a Lindblad master equation for the 
reduced 
density matrix $\varrho$ of a quantum system\footnote{Here, $\hbar = 1$ and we 
apply this convention throughout the paper.} \cite{Breuer02,Schaller14}
\begin{equation}\label{eq:LB}
 \frac{\D\varrho}{\D t} = 
 -\II [H,\varrho] + \sum_{\vec{n}} L_{\vec{n}}  \varrho L_{\vec{n}}^\dagger - 
 \frac{1}{2} \{L_{\vec{n}}^\dagger L_{\vec{n}},\varrho\} = \mathcal{L}(\varrho)\,.
\end{equation}
\textcolor{black}{Here, $H$ is the quantum Hamiltonian of the system describing
the \textit{coherent} part of the dynamics which accounts for wave-like phenomena such as superposition
and interference to take place. $L_n$ are
certain quantum jump operators, each of which introduces an \textit{incoherent}
stochastic process to the dynamics of the quantum system. Thus, the above Lindblad dynamics
is suitable to describe the interplay between classical hopping and quantum coherent evolution~\cite{Whit10}.}

The generic solution of Eq.~\eqref{eq:LB} with the initial density matrix
$\varrho(0)$ can be written in terms of the time-independent superoperator
$\mathcal{L}$ as 
\begin{align}
 \varrho(t) = e^{\mathcal{L} t} \varrho(0)\,.
\end{align}

In the following sections, we employ a set of rules~\cite{Whit10} to include CRWs 
as stochastic background to QWs and we shall see that each classically
allowed transition will result in a dissipative contribution to the superoperator.

We proceed as follows. In Secs.~\ref{sec:crw} and \ref{sec:qrws}, we specify the 
types of CRWs and QWs on networks that we study in this paper. These can be
seen as cornerstones of the outlined theory.
In Sec.~\ref{sec:qsws2}, we introduce an interpolation scheme between these
cornerstones and, in this way, define a quantum-to-classical random 
walk (QCW).\footnote{We use the term ``quantum-to-classical'' 
random walk to indicate that we consider a subset of QSWs that interpolates between CRWs and QWs.}
Finally, in Sec.~\ref{sec:rqsws}, we introduce stochastic 
resetting by means of a suitable quantum jump process.

%
%
%
\subsection{Classical random walks on networks}
\label{sec:crw}
\noindent
We describe the evolution of a CRW on a network $G$ in terms of the 
probabilities $p_i(t)$ of observing a walker on node $i$ at time $t$.
These probabilities form a normalized probability vector
\begin{equation}
 \mathbf{p}(t) = \left(p_1(t),\dots,p_N(t)\right), \quad \sum_{i=1}^N p_i(t) =1 \, .
\end{equation}

In the time interval $[t, t+\Delta t]$, conservation of probability 
implies that the local probabilities can only flow in and out of nodes. In 
particular, if $\Delta t$ is small enough, the net-influx to node $i$ originates from the immediate neighborhood of node $i$.
The evolution of the probability $p_i(t)$ is thus described by the following
rate equation
\begin{align}
\begin{split}
p_i(t+\Delta t)-p_i(t)&=
%
%
-\Delta t \bigg( p_i(t)-\sum_{j=1}^N \frac{A_{ i j}}{k_j} p_j(t)\bigg) \, .
\end{split}
\label{eq:class_ran_walk1}
\end{align}
The first term on the right-hand-side corresponds to the outward flow from
node $i$ to its surroundings
and the second term describes the inflow to node $i$. 
We may rewrite Eq.~\eqref{eq:class_ran_walk1} in matrix form
\begin{align}
\begin{split}
\frac{\mathbf{p}(t+\Delta t)-\mathbf{p}(t)}{\Delta t}
=-  \mathbb{L} D^{-1} \mathbf{p}(t) \, ,
\end{split}
\label{eq:class_ran_walk2}
\end{align}
with the graph Laplacian $\mathbb{L}$ and the degree matrix $D$ (see Eqs.~(\ref{eq:D}) and~(\ref{eq:L})). In the limit $\Delta t \to 0$, this rate equation becomes a master equation 
in differential form
\begin{align}\label{eq:crw}
 \frac{\D}{\D t} \mathbf{p}(t) &= -H_{\rm c} \mathbf{p}(t),
\quad
  H_{\rm c} = \mathbb{L} D^{-1}  
.
\end{align}
Here, we introduced the classical Hamiltonian $H_{\rm c}$ as generator of time translation
of the probability distribution.
Equation~\eqref{eq:crw} is formally solved with the time evolution 
operator $S(t)=e^{-H_{\mathrm{c}} t}$ which allows us 
to write the probability distribution at time $t$, that originated from an initial distribution $\mathbf{p}_0$, as 
\begin{equation}
\mathbf{p}(t) = S(t) \mathbf{p}_0 \, .
\label{eq:class_ran_walk3}
\end{equation}
For connected networks, where one can reach any node from any other node, and sufficiently long times, the CRW approaches a 
stationary probability distribution $\mathbf{p}^*$ such that
$p_i^* = \sum_{j=1}^N \frac{A_{ij}}{k_j} p_j^*$. This allows us to fully determine the 
stationary probability distribution as\footnote{It is straightforward to check that $\bf p^*$ is 
indeed the steady state by applying the classical Hamiltonian to Eq.~\eqref{eq:cl_stat}.
}
\begin{equation}
p_i^* = \frac{k_i}{\sum_{i=1}^N k_i}\, .
\label{eq:cl_stat}
\end{equation}
\textcolor{black}{In the following sections, we will compare CRWs and QWs in terms of the 
probability $p_k^\prime$ that any \textcolor{black}{of the $n_k$ nodes} with degree $k$ is occupied. Note that $p_k^\prime$ satisfies  
\begin{equation}
\sum_{k }n_k p_k^\prime = 1 \,,
\label{eq:pk}
\end{equation}
whereas we have $\sum_{i=1}^N p_i^\ast = 1$ in the node-centered formulation of the occupation probability.}
%
%
%
\subsection{Quantum walks on networks}
\label{sec:qrws}
\begin{figure*}[t]
 \includegraphics[width=.3\textwidth]{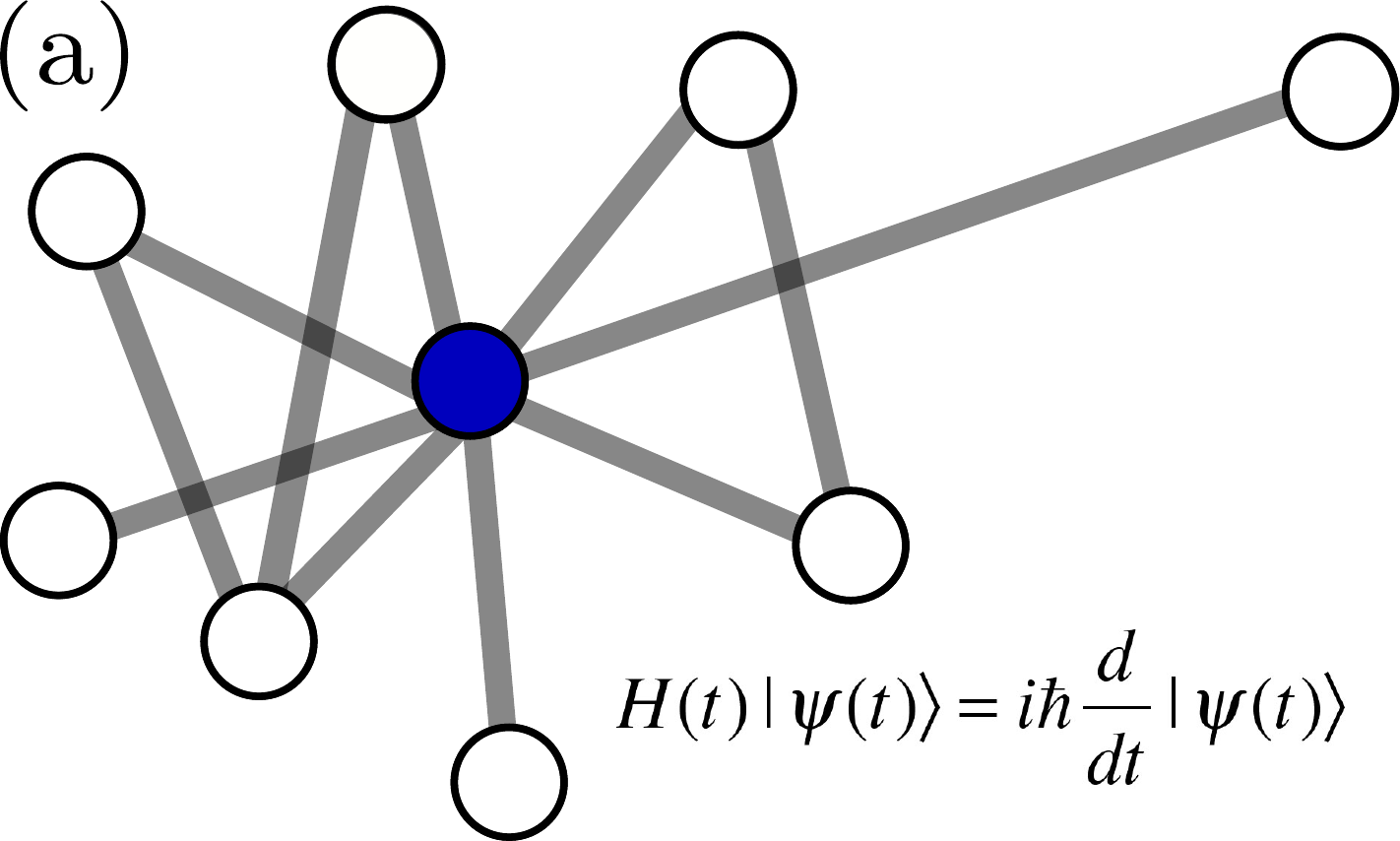}\qquad
 \includegraphics[width=.3\textwidth]{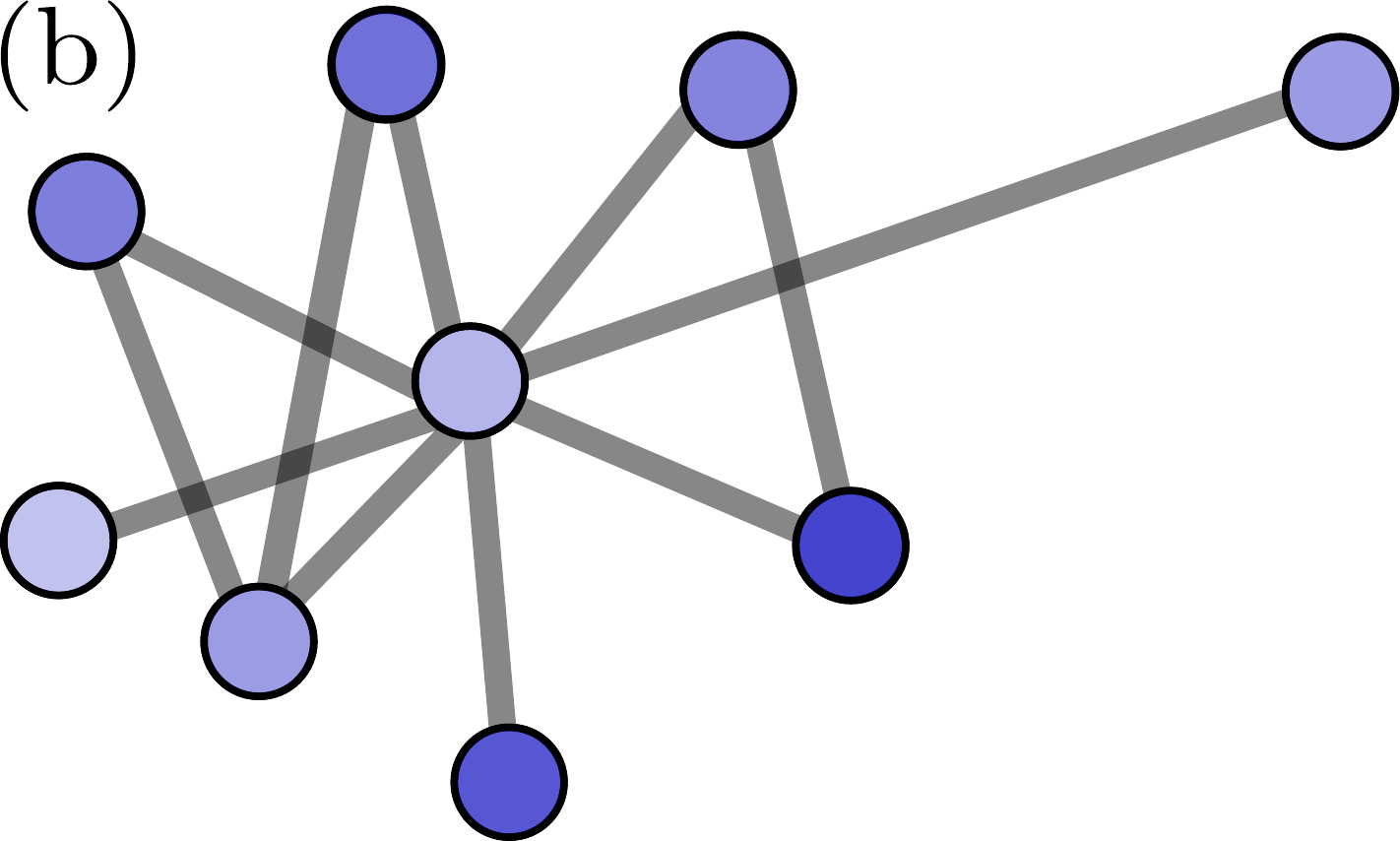}\qquad
 \includegraphics[width=.3\textwidth]{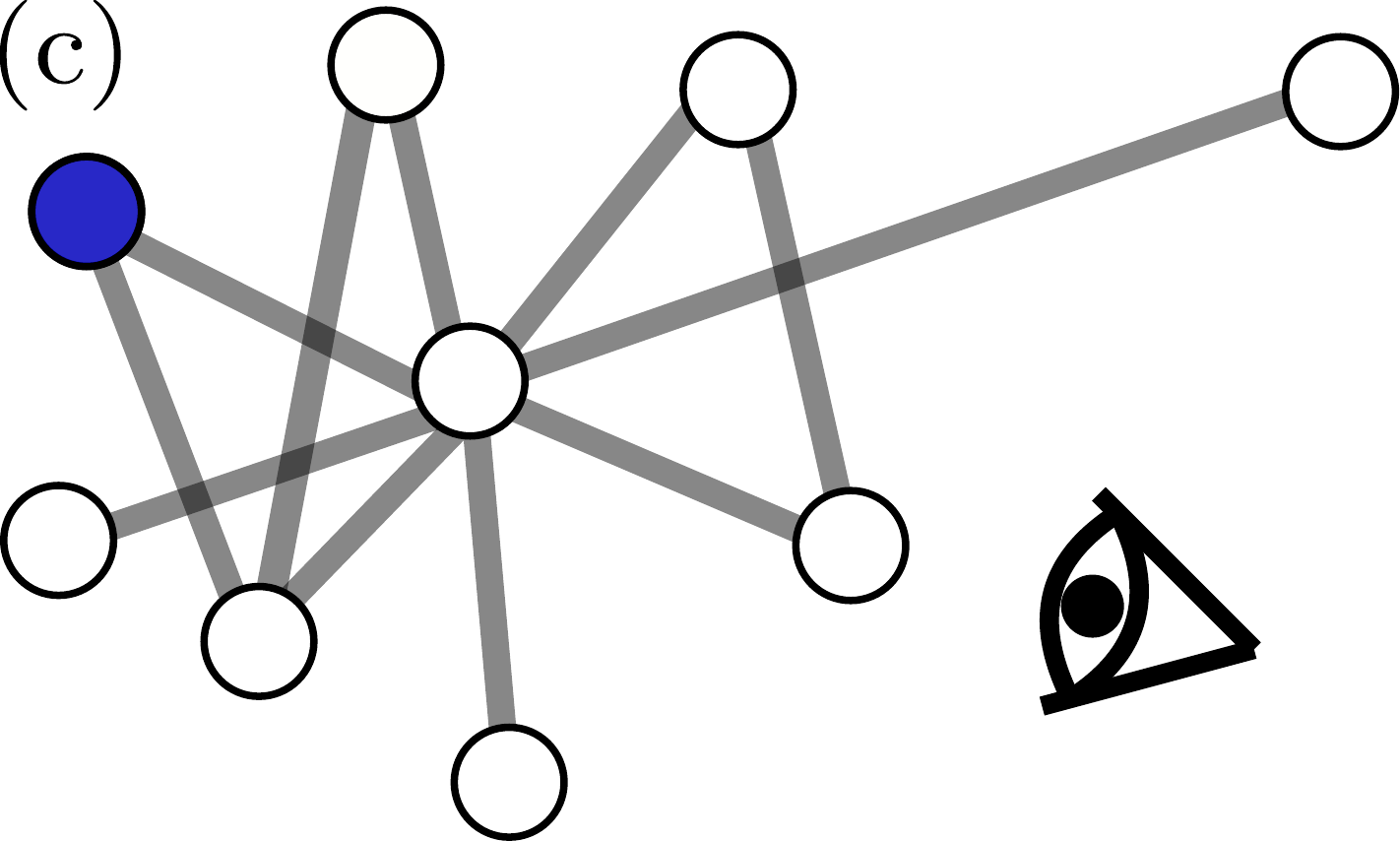}
 \caption{{\bf Schematic of a quantum walk.} 
 A quantum walker starts from a certain initial state in panel (a). Here the initial
 state is fully localized on a certain node for illustration purposes. The dynamics of
 the quantum walker is governed by the Schr\"odinger equation. After some time, the wave 
 function of the walker will be spread over the network as depicted in panel (b)
 and there will be quantum superpositions. Knowing the initial state, both the 
 Hamiltonian that induces the dynamics and the waiting time uniquely 
 determine this state. Observing (or measuring) the quantum walker will
 result in a collapse of the wave function as shown in panel (c) and there 
 is no way of knowing to which state the wave function collapses. This 
 introduces stochasticity to QWs.
 }
 \label{fig:qw}
\end{figure*}

\noindent
A quantum walker is described by its wave function $\ket{\phi}\in\mathbb{C}^N$ rather than 
a probability distribution. The wave function propagates according to the 
Schr\"odinger equation
\begin{equation}
 \partial_t \ket{\phi} = - \II H_{\rm q} \ket{\phi}
\label{eq:qw}
\end{equation}
with some quantum Hamiltonian $H_{\rm q}$.
This formulation of QWs is generally referred to as continuous time quantum walk~\cite{Ven12}.
We note that a QW, as defined in Eq.~\eqref{eq:qw}, is not inherently stochastic since the Schr\"odinger equation itself is  deterministic.
Stochasticity in QWs rather stems from measurements that are applied to the quantum system and lead to
a collapse of the wave-function~\cite{kollar2012asymptotic} as illustrated in Fig.~\ref{fig:qw}.
In analogy to a CRW (see Eq.~\eqref{eq:class_ran_walk3}), the quantum Hamiltonian generates a quantum-time-evolution operator
\begin{equation}
U(t)=e^{-\II H_\mathrm{q} t}
\end{equation}  
for a QW on $G$.
The choice of $H_{\rm q}$ thus determines the behavior of the QW.
We follow Ref.~\cite{faccin2013degree} and choose the symmetric and normalized graph Laplacian
\begin{equation}
H_\mathrm{q}=D^{-1/2} \mathbb{L} D^{-1/2}  
\label{eq:Hc}
\end{equation}
as hermitian quantum Hamiltonian.\footnote{Note that the classical Hamiltonian is
not necessarily Hermitian since in general $[A,D^{-1}] \neq 0$ (for a lattice 
we though have $[A,D^{-1}] = 0$ since $D\propto\mathbb{1}$).
}
QWs that are based on the Hamiltonian in Eq.~\eqref{eq:Hc} have the appealing property that the average probability to find the 
quantum walker on a certain node will be the same as in the 
classical case if the system is in the ground state
(see Sec.~\ref{sec:crw})~\cite{faccin2013degree,Biamonte2019}. 
%
%
%

In order to prepare the
ground for more general quantum stochastic walks,
we replace the wave function $\ket{\phi}$
by the density matrix $\varrho = \sum_{ij} \varrho_{ij} \ket{i}\bra{j}$ and the 
Schr\"odinger equation (see~Eq.~\eqref{eq:qw}) by the equivalent von-Neumann equation
\begin{equation}\label{eq:qw2}
 \frac{\D \varrho}{\D t} = -\II \left[ H_{\rm q}, \varrho \right]\, .
\end{equation}
This formulation is able to account for statistical mixtures
of wave functions and can be easily extended to open quantum systems as needed for
classical-quantum mixtures of random walks.
For the QW that results from Eqs.~(\ref{eq:Hc}) and~(\ref{eq:qw2}), the long-time average probability of being on node $i$
is given by
\begin{equation}\label{eq:cesaro}
 q_i^* = \lim_{T\rightarrow\infty}\frac{1}{T} \int_0^T  \ \bra{i}
 \varrho (t) \ket{i} \D t\, .
\end{equation}
\textcolor{black}{Similar to the classical case, we denote by $q_k'$ the probability that any node with 
degree $k$ is populated.
As in Eq.~\eqref{eq:pk}, the probability $q_{k}^\prime$ satisfies
\begin{equation}
\sum_{k} n_k q_k^\prime = 1\, .
\label{eq:qk}
\end{equation}
}
%
%
%
\subsection{Quantum-to-classical stochastic walks}
\label{sec:qsws2}
\noindent
Having outlined frameworks for the treatment of CRWs and QWs on networks, we now proceed and introduce a generalized quantum stochastic walk that interpolates between
these walks.
We recall that any such walk is described by the Lindblad master equation~(\ref{eq:LB})
and introduce a dimensionless interpolation parameter $\epsilon$ in such a way that we recover the QW of Eq.~(\ref{eq:Hc}) for $\epsilon = 0$.
The choice
\begin{equation}
 H = (1-\epsilon) H_{\rm q}\, , \qquad L_n \propto \sqrt{\epsilon}\, ,\quad \forall n
\end{equation}
reduces the Lindblad master equation to the von-Neumann equation of the QW (see~Eq.~(\ref{eq:qw2})), in the limit $\epsilon \to 0$. 

In order to also account for CRW dynamics, we need 
to choose specific dissipative processes by specifying the quantum jump operators
$L_n$. It has been 
shown that if the quantum jump operators satisfy the relation~\cite{Whit10}
\begin{align}\label{eq:cdt1}
\begin{split}
 \sum_{\vec{n}}\bigg( \delta_{ab}&\bra{a}  L_{\vec{n}}^\dagger L_{\vec{n}} \ket{a} \\
 &+ \sum_{{\vec{m}}} \bra{a} L_{\vec{n}} \ket{b} \bra{b}L_{\vec{m}} \ket{a} \bigg)
 = - \bra{a}H_{\rm c}\ket{b} ,
 \end{split}
\end{align}
the classical processes induced by the classical Hamiltonian
are encompassed in the dynamics of the generalized quantum system.
In order to proceed, we choose the set of 
quantum jump operators\footnote{The explicit choice of the quantum jump operators is not unique.}
\begin{equation}\label{eq:cdt2}
 L_{\vec{n}} = L_{nm} = \sqrt{\epsilon\gamma_{nm}} \ket{n}\bra{m}, \,
 \vec{n}= (n,m)\in E \, .
\end{equation}
$L_{nm}$ induces jumps from node $n$ to node $m$ and satisfies
\begin{equation}\label{eq:cdt3}
 L_{nm}^\dagger L_{nm} =  \epsilon \gamma_{nm}\ket{m}\bra{m}.
\end{equation}
The parameters $\gamma_{nm} \in \mathbb{R}$ are the so-called 
{\it damping constants}. We use Eqs.~\eqref{eq:cdt1} and \eqref{eq:cdt2}
to determine the damping constants, viz.
\begin{equation}
 \gamma_{nm} = -\delta_{nm} +  \frac{A_{nm}}{k_m}
= -\bra{n}H_{\rm c}\ket{m}\, .
\end{equation}
It is important to distinguish between diagonal and off-diagonal contributions. 
Based on Eq.~(\ref{eq:crw}), it is clear that the off-diagonal elements of the classical 
Hamiltonian are negative, rendering the damping constants $\gamma_{nm}$ positive for 
$n\neq m$. For $n=m$, the entries of the classical Hamiltonian are equal to $1$.
Consequently, the parameter $\gamma_{nn}<0$ and thus
\begin{equation}
 L_{nn} = \II \sqrt{\epsilon} \ket{n}\bra{n}\, .
\end{equation}
The full dynamics of a QCW is thus given by the 
Lindblad master equation
%
%
%
\begin{align}\label{eq:qsw}
\begin{split}
 \frac{\D \varrho}{\D t} &= -\II \left[ (1-\epsilon) H_{\rm q}, \varrho \right]\\
 &+\epsilon\sum_{nm}\bra{n}H_{\rm c}\ket{m}\left[
 \varrho_{mm} \proj{n}{n}-\frac{1}{2}\{ \proj{m}{m}, \rho\}\right]
 \end{split}
 \end{align}
%
%
%
Based on this expression, it is possible to explicitly show that this QCW recovers
the CRW (see Eq.~\eqref{eq:crw}) in the limit $\epsilon\to 1$, see
App.~\ref{app:eps1} for further details.

\textcolor{black}{Despite the {\it a priori} phenomenological character of this approach 
to incorporate coherent and incoherent dynamical aspects, Eq.~(\ref{eq:qsw}) can 
be considered as \textcolor{black}{generic} dynamics for a walker on a given graph structure.
The parameter $\epsilon$ then indicates the competition between coherent and incoherent
processes between neighboring nodes. This is most evident from an algorithmic point of 
view\footnote{See Sec.~\ref{sec:numerical_recp} for a detailed description of the unfolding of the quantum master equation 
as a stochastic Schr\"odinger equation.}:
Imagine a discrete time step $\Delta t$ at the beginning of which a localized walker undergoes a unitary dynamics, 
meaning that the wave function spreads across the graph, see Fig.~\ref{fig:qw} (a) and (b).
In the next time step, there is
\begin{itemize}
\item
a certain probability proportional to $\epsilon \Delta t$ that an incoherent transition occurs from any 
node that the wave function occupies to a corresponding adjacent site.
\item 
a certain probability that the wave function spreads further in the neighborhood of already
occupied sites.
\end{itemize}
} 
\noindent
\textcolor{black}{In this manner, coherent and incoherent processes compete on the same graph in our hybrid quantum-to-classical walk formulation~(\ref{eq:qsw}). Similar formulations have already proven to be of great use, e.g.~in the
design of quantum versions of the Page Rank~\cite{Buri12}, for graph isomorphism problems~\cite{Bru16}, dissipative quantum computing algorithms~\cite{Attal12,Sin12}, and decision-making~\cite{Mart16}.}

As the dissipator in Eq.~(\ref{eq:qsw}) is well defined by the classical 
and quantum Hamiltonian and the parameter $\epsilon$, we introduce the 
short-hand notation in terms of the superoperator
\begin{align}
 \frac{\D \varrho}{\D t}  = \mathcal{L}^{(\epsilon)}(\rho) \ .
\end{align}
The QCW defined by Eq.~(\ref{eq:qsw}) linearly interpolates between the CRWs and QWs
that we defined in the preceding sections. This means that classical and quantum dynamics have been induced on the {\it same network} and that each 
link of this network is capable of hosting a classical and a quantum hopping process.
One way to look at this is that finite thermal excitations in the system may introduce classical hopping on the quantum graph.
This setup can be readily altered by, for instance, defining separate quantum and classical layers. We leave these directions for future
works and focus instead on the dynamics induced by Eq.~(\ref{eq:qsw}).

\subsection{Reset quantum stochastic walks on networks}\label{sec:rqsws}
\noindent
We describe stochastic resets 
by an additional dissipative contribution in the evolution of QCWs (see Eq.~\eqref{eq:qsw})~\cite{rose2018spectral}.
The modified Lindblad master equation 
including a reset process with a certain rate $r$ to the initial state $\varrho(0)$ reads
\cite[Eq.~(59)]{rose2018spectral}
\begin{align}\label{eq:reset}
 \partial_t \varrho &= \mathcal{L}^{(\epsilon)}(\varrho)
%
 +r \varrho(0) - r \varrho 
 \equiv \mathcal{L}^{(\epsilon)}_{r}(\varrho)\, .
\end{align}
In the case of a pure reset state, the reset density matrix may be written
as a projector $\varrho(0) = \ket{\phi(0)}\bra{\phi(0)}$.
It turns out that the stationary state of the reset dynamics $\varrho_r^*$
may be written explicitly in terms of left and right eigenmatrices and eigenvalues
of the system without reset
\begin{equation}
 \mathcal{L}^{(\epsilon)}_0(\mathfrak{r}_n^{(\epsilon)}) =  \lambda_n^{(\epsilon)} \mathfrak{r}_n^{(\epsilon)}, \qquad
 \big(\mathcal{L}^{(\epsilon)}_0\big)^\dagger(\mathfrak{l}_n^{(\epsilon)}) =  \bar{\lambda}_n^{(\epsilon)} \mathfrak{l}_n^{(\epsilon)}\  .
\end{equation}
The stationary state then reads~\cite{rose2018spectral} 
\begin{equation}\label{eq:statopen}
 \big(\varrho_{r}^{(\epsilon)}\big)^* = \big(\varrho_{r=0}^{(\epsilon)}\big)^* + r \sum_{n=2}^{N^2}
\frac{\bra{\phi(0)} \big(\mathfrak{l}_n^{(\epsilon)}\big)^\dagger \ket{\phi(0)}}{\lambda_n^{(\epsilon)} - r}
\mathfrak{r}_n^{(\epsilon)}.
\end{equation}
This holds for all $0 < \epsilon \leq 1$ and reset rates $r\geq 0$. 
In the case of a QW ($\epsilon = 0$), there is no stationary state without reset and Eq.~\eqref{eq:statopen} reduces to~\cite{rose2018spectral}
\begin{align}\label{eq:statclosed}
  \big(\varrho_r^{(1)}\big)^* = E \Lambda_r E^\dagger,
\end{align}
where $E$ is the matrix of eigenvectors of the quantum Hamiltonian
$H_{\rm q} E = E \Lambda$ with $\Lambda_{ij} = \lambda_i\delta_{ij}$ and the elements of $\Lambda_r$ are
\begin{equation}
 (\Lambda_r)_{ij} = r \frac{ \sandwich{\phi(0)}{e_j}\sandwich{e_i}{\phi(0)}}{
 r + \II (\lambda_i - \lambda_j)}\, .
\end{equation}
Equations~\eqref{eq:statopen} and \eqref{eq:statclosed} are remarkable since they express the stationary state of the reset dynamics in terms 
of the unperturbed system without reset (i.e., for $r=0$). This means that, e.g.~for a QW where $\epsilon = 0$, the stationary state for $r>0$ 
is found analytically by diagonalizing the quantum Hamiltonian.
Furthermore, for a general reset-QCW, the closed form allows us to determine the steady-state 
probability for the walker to be on node $\ell$, viz.
\begin{align}
\begin{split}
\big(q_{r}^{(\epsilon)}\big)_\ell^*
&= \big(q_0^{(\epsilon)}\big)_\ell^* + r\sum_{n=2}^{N^2}  
\frac{\bra{\phi(0)} \mathfrak{l}_n^{(\epsilon)\dagger} \ket{\phi(0)}\bra{\ell} \mathfrak{r}_n^{(\epsilon)}\ket{\ell}}{\lambda_n^{(\epsilon)} - r} 
  .
\label{eq:statsolq}
\end{split}
 \end{align}
\textcolor{black}{In the special case of a QW ($\epsilon = 0$), this formula reduces 
to~\cite{rose2018spectral}
\begin{widetext}
\begin{align}
\begin{split}
\big(q_{r}^{(0)}\big)_\ell^*
%
&= 
 \sum_{j,k} \frac{r \sandwich{\phi(0)}{e_k}\sandwich{e_{j}}{\phi(0)}}{r+
 \II (\lambda_k-\lambda_{j})}  \sandwich{\ell}{e_{j}} \sandwich{e_{k}}{\ell}\\
 &=\int_0^\infty r e^{-r \tau} \sum_{j,k}  e^{-\II (\lambda_j-\lambda_k)\tau} \sandwich{\phi(0)}{e_k}\sandwich{e_{j}}{\phi(0)} \sandwich{\ell}{e_{j}} \sandwich{e_{k}}{\ell} \, \mathrm{d}\tau ,
\end{split}
\label{eq:statsolqQW}
 \end{align}
\end{widetext}
}
\noindent
where $\lambda_k$ and $\ket{e_k}$ are the eigenvalues and the eigenvectors 
of the quantum Hamiltonian $H_{\rm q}$. That is,
\begin{equation}
 H_{\mathrm{q}} \ket{e_k} = \lambda_k \ket{e_k}.
\end{equation}

For a CRW (or QCW with $\epsilon=1$), it is also possible to directly determine the 
stationary probability distribution ${\bf p}^*(r)$. In 
an infinitesimal time interval $[t,t+\mathrm{d}t]$, the classical walk starts 
from its initial state $\mathbf{p}_0$ with probability $r \mathrm{d}t$. The 
corresponding reset times $\tau$ are exponentially distributed with 
probability-density function $\varphi(\tau)=r e^{-r \tau}$. 
\textcolor{black}{If the reset 
rate is finite (i.e., $r>0$), we find the corresponding stationary
distribution (see App.~\ref{app:laplace} for further details)
%
%
%
\begin{align}
\begin{split}
\mathbf{p}^*(r)
&= \int_0^\infty r e^{-r \tau} e^{-H_{\mathrm{c}}\tau}\mathbf{p}_0\, 
 \mathrm{d}\tau \\
 &\textcolor{black}{=\int_0^\infty r e^{-r \tau} \sum_n e^{-\lambda_n \tau} (\mathbf{p}_0)_n \ket{n}} \,\mathrm{d}\tau\\
&=r \left[\mathbb{1}(1+r)-AD^{-1}\right]^{-1}\mathbf{p}_0 \ .
\label{eq:pr_stat}
\end{split}
\end{align}
Note the similarity in the mathematical structure of Eqs.~\eqref{eq:statsolqQW} and \eqref{eq:pr_stat}. One marked difference in $\big(q_{r}^{(0)}\big)_\ell^*$ is the appearance of product states that result from the mixing of wave-function components.}
\begin{figure*}[t]
\centering
\includegraphics[width=0.3\textwidth]{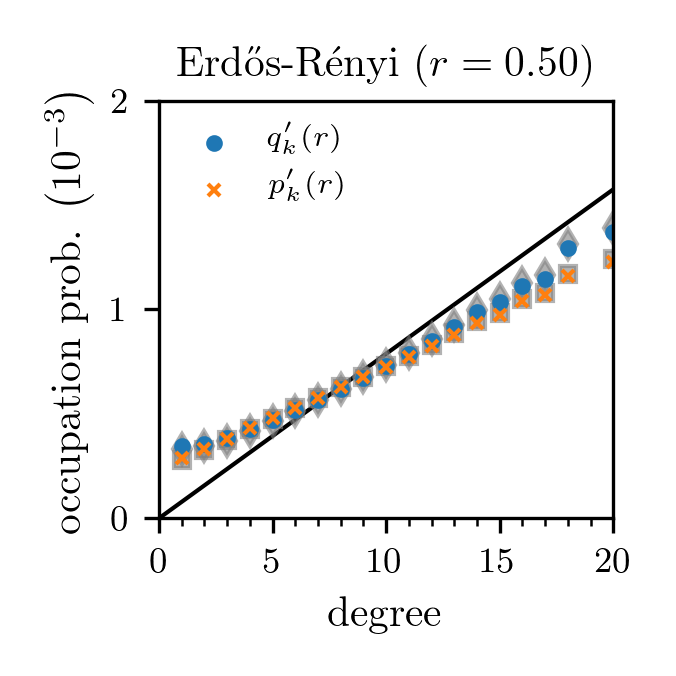}
\includegraphics[width=0.3\textwidth]{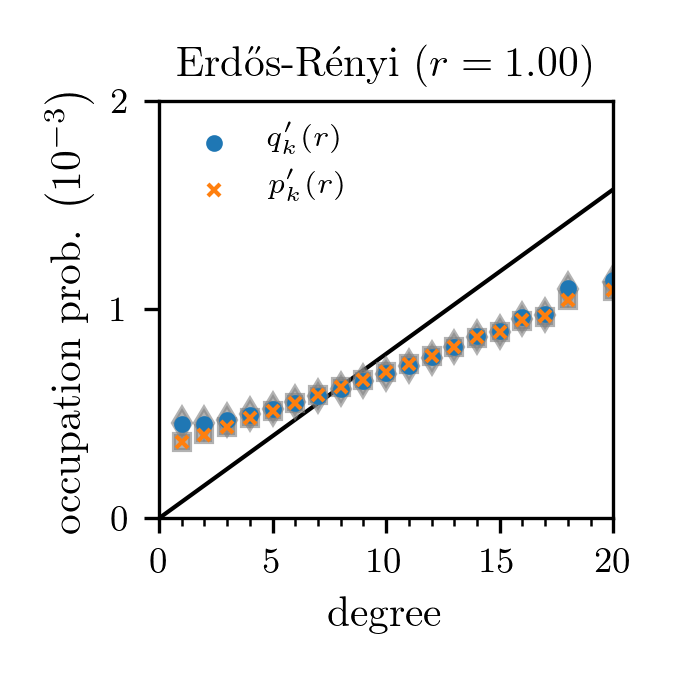}
\includegraphics[width=0.3\textwidth]{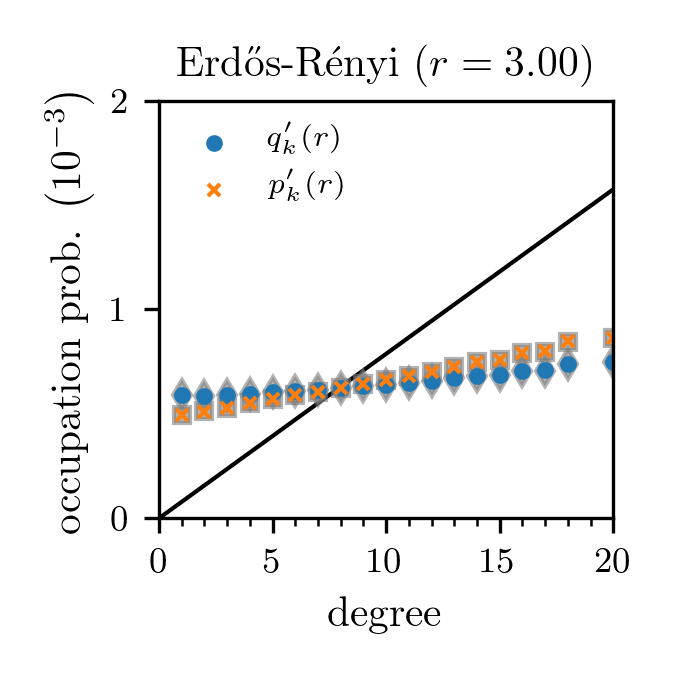}\\[-.3cm]
\includegraphics[width=0.3\textwidth]{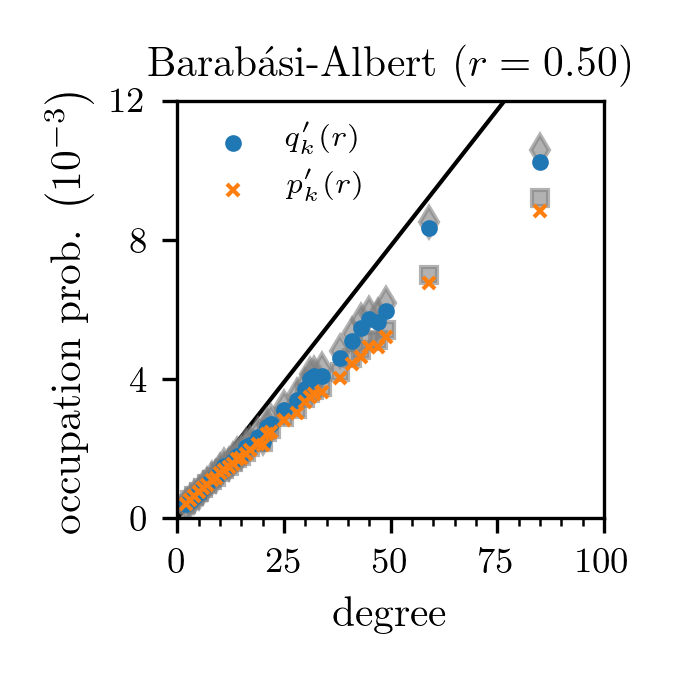}
\includegraphics[width=0.3\textwidth]{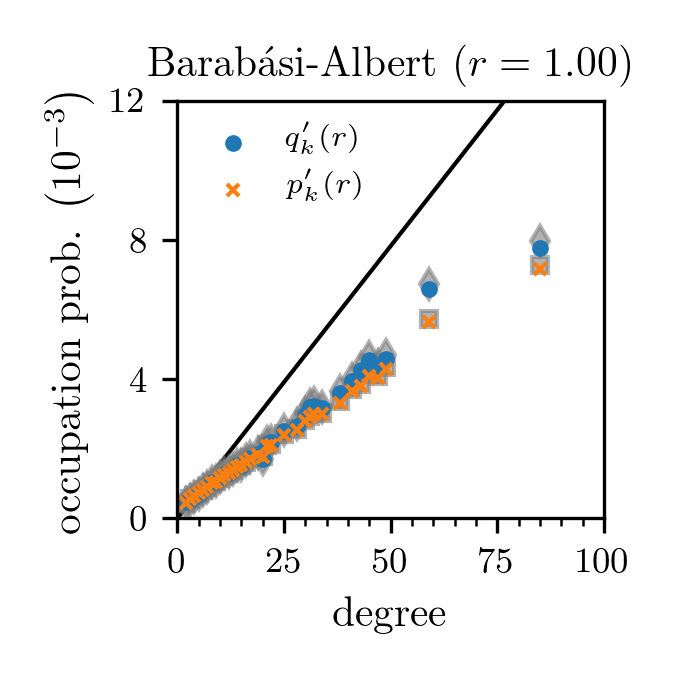}
\includegraphics[width=0.3\textwidth]{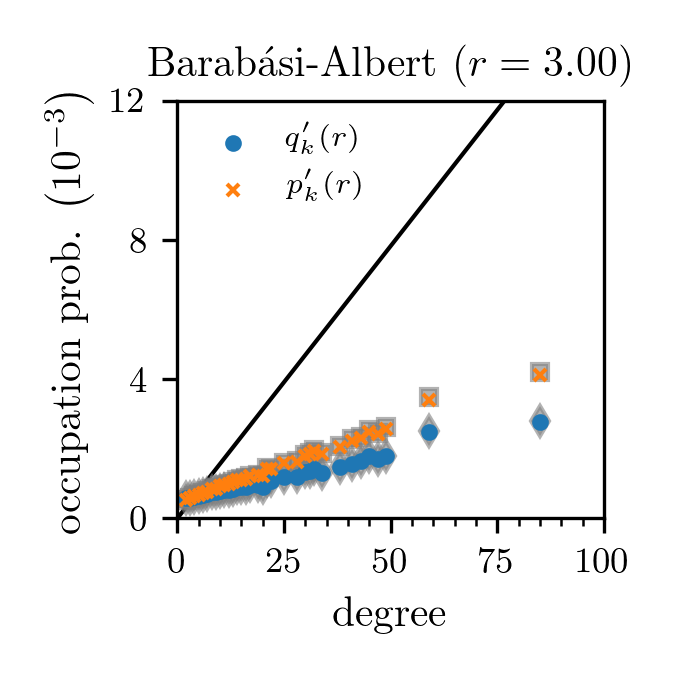}\\[-.3cm]
\includegraphics[width=0.3\textwidth]{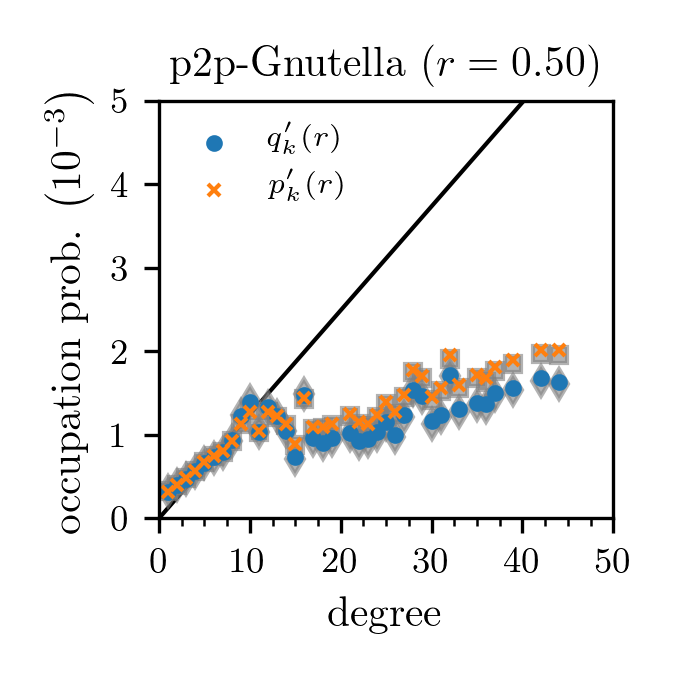}
\includegraphics[width=0.3\textwidth]{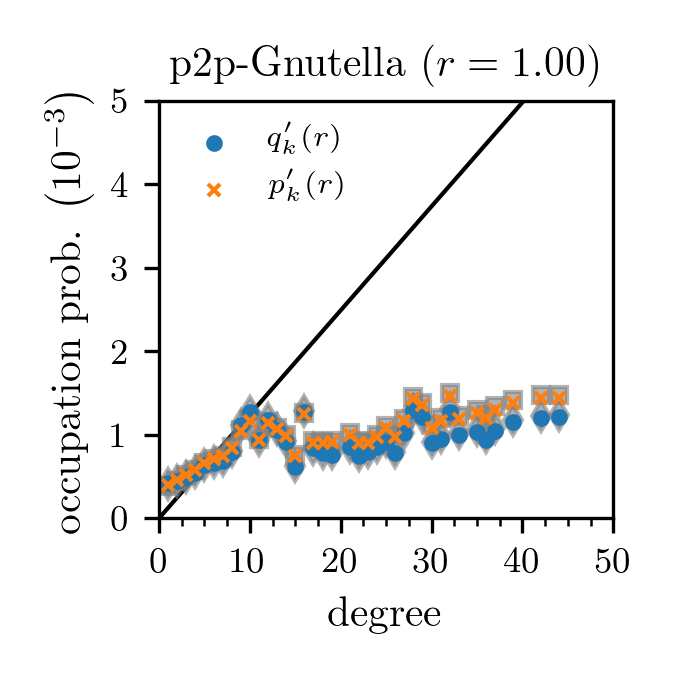}
\includegraphics[width=0.3\textwidth]{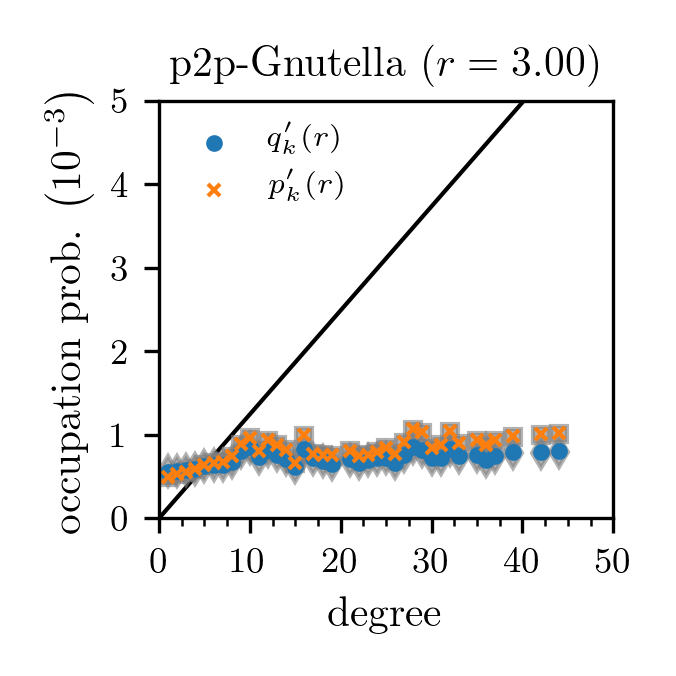}
\caption{\textbf{Occupation probability for different networks and reset rates.} 
For different networks (row) and reset rates $r$ (column), we show the probabilities 
$p^\prime(r)$ and $q^\prime(r)$ (see Eqs.~\eqref{eq:pk} and \eqref{eq:qk}) that a node of degree 
$k$ is occupied by a classical and quantum random walker, respectively. As 
initial conditions and reset states, we use a uniform distribution over all nodes. Our numerical results are based on solutions of 
Eqs.~\eqref{eq:class_ran_walk2} and \eqref{eq:crank_nicholson}. The networks we 
consider have $N=1600$ nodes. The shown data points are averages over $2.5\times 10^5$ 
samples. We use grey markers to indicate the solutions of the analytic results
Eqs.~\eqref{eq:statsolqQW} and \eqref{eq:pr_stat}. The black solid line is the occupation probability of a CRW for $r=0$
as reference.
}
\label{fig:comparison1}
\end{figure*}
%
%
%

%
%
%
\section{Numerical Recipes}
\label{sec:numerical_recp}
\noindent
In order to efficiently model classical and quantum random walks on large networks, we simulate the Lindblad
master equation as {\it piecewise deterministic process}. The commonly used 
(but not unique) unfolding 
of the master equation~(\ref{eq:LB}) in terms of a stochastic Schr\"odinger equation reads~\cite{Breuer02,Schaller14}
\begin{align}\label{eq:sse}
 \begin{split}
 \ket{\D \phi}  =  -\II  H_{\rm eff}\ket{\phi} \D t 
 + \sum_n \bigg[\frac{L_n \ket{\phi}}{\sqrt{\bra{\phi}L_n^\dagger L_n
 \ket{\phi}}} -\ket{\phi}\bigg] \D N_n \, .
 \end{split}
\end{align}
Here $\ket{\phi}$ is the wave function of a certain quantum trajectory and the Poisson increment $\D N_n$ 
describes a noisy contribution that is generated by the physical process belonging to the quantum jump operator
$L_n$. In each simulation step, the system either performs a time evolution according to the effective (non-hermitian) Hamiltonian
\begin{equation}
 H_{\rm eff} =  H -\frac{\II}{2} \left[\sum_n L_n^\dagger L_n  -
\bra{\phi} L_n^\dagger L_n \ket{\phi}\right]
\end{equation}
or otherwise an instantaneous quantum jump $\D N_{n}$ occurs. Quantum jumps $\D N_{n}$ satisfy
\begin{align}
 \D N_{n} \D N_{m}= \delta_{nm} \D N_{n}, \ \,
 \left< \D N_{n} \right> =  \bra{\phi}L_n^\dagger L_n \ket{\phi} \D t.
\end{align}
The probability that a jump process occurs in the time step $\D t$ is
\begin{align}\label{eq:Pj}
 P_{\rm j} = \sum_{n}  \D t \bra{\phi}L_n^\dagger L_n \ket{\phi}.
\end{align}
The average over independently sampled quantum trajectories (or respectively 
the long-time limit of a single trajectory) allows us to determine the results of the 
Lindblad master equation.
\subsection{Stochastic Schr\"odinger equation for the reset dissipator}
\begin{figure*}[t]
\centering
\includegraphics[width=\textwidth]{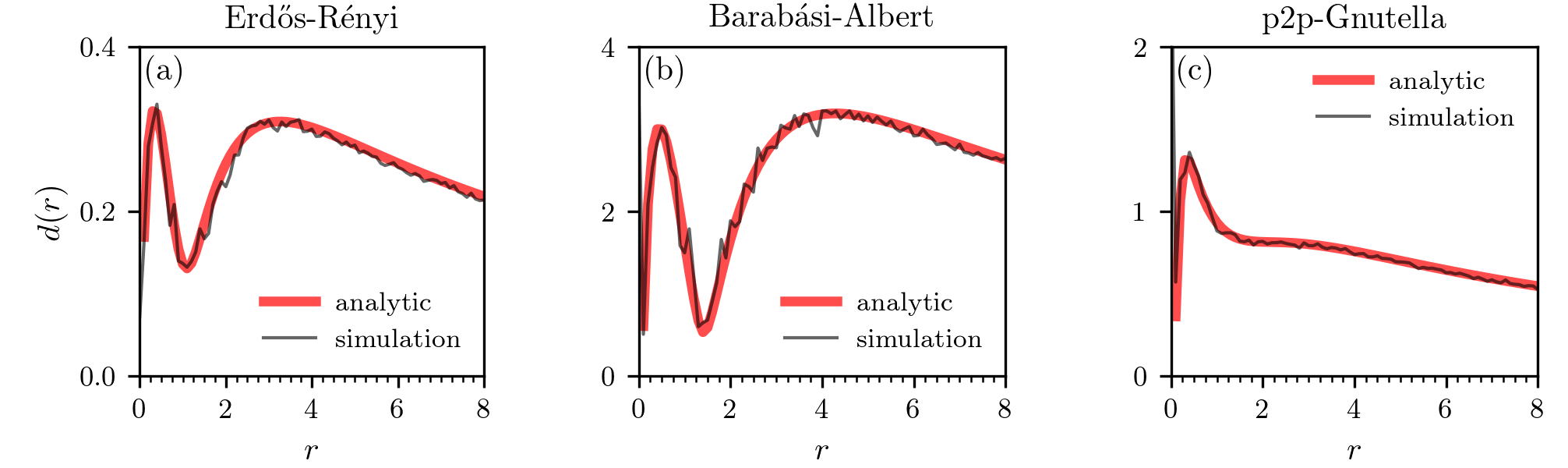}
\caption{\textbf{Difference between the classical and quantum occupation probability for different networks and reset rates.} 
For different networks, we show the distance $d(r)$ (see Eq.~(\ref{eq:d})). As initial condition, we use a uniform distribution over all nodes. Our numerical results are based on solutions of 
Eqs.~\eqref{eq:class_ran_walk2} and \eqref{eq:crank_nicholson}. The networks we 
consider have $N=1600$ nodes. The shown data points are averages over $2.5\times 10^5$ 
samples.
}
\label{fig:comparison2}
\end{figure*}
\noindent
Here we discuss the numerical implementation of a quantum stochastic reset process for an otherwise unitary quantum walker. The 
inclusion of further dissipative processes is straightforward as the dissipative processes are additive in the Lindblad master equation
(\ref{eq:LB}). We shall exploit this fact in the next sections to write down the stochastic Schr\"odinger equation for 
a QCWs.

The quantum jump operators for the reset process presented in Sec.~\ref{sec:rqsws} are~\cite{rose2018spectral}
\begin{equation}
 J_{n}^r = \sqrt{r} \ket{\phi(0)}\bra{n}.
\end{equation}
From Eq.~(\ref{eq:sse}) we obtain the corresponding stochastic 
Schr\"odinger equation 
\begin{align}
\begin{split}
 \ket{\D \phi} 
=  r\sum_{n\in V} \left[\frac{\sandwich{n}{\phi}}{\left|
 \sandwich{\phi}{n}\right|}\ket{\phi(0)} -\ket{\phi}\right] \D N_n
 -\II H \ket{\phi} \D t
 \end{split}
\end{align}
with the quantum jump probability
\begin{equation}
 P_{\rm j} = r \D t\, .
\end{equation}
We see that the quantum jump probability corresponds to the reset 
rate and does not depend on the current quantum state, as it should
be for a stochastic reset process.
Furthermore, the only difference to a brute-force reset is a global 
phase that keeps information about the pre-reset state. Since this phase is global, we neglect it in the remainder and write 
\begin{align}
 \ket{\D \phi}   
=  &-\II H \ket{\phi} \D t
 + r\sum_{n\in V} \bigg[\ket{\phi(0)} -\ket{\phi}\bigg] \D N_n \ .
%
\end{align}
This stochastic Schr\"odinger equation then yields the following simple
stochastic rule for the time evolution of a unitary process with 
dissipative stochastic resets
\begin{align}
\begin{split}
\ket{\phi(t+\mathrm{d}t)} =
&\left[1-\II H_\mathrm{q} \mathrm{d}t \right]\ket{\phi(t)} \Theta(z-r\, \D t)\\[.25cm]
&+\ket{\phi(0)} \Theta(r\,\mathrm{d}t-z)
\end{split}
\label{eq:qwalk_reset}
\end{align}
with a uniformly distributed random number $z\in[0,1]$ and the Heaviside step function $\Theta(x)$, which is 1 for $x \geq 0$ and 0 otherwise. 
This description of quantum reset processes has been also used in previous
studies~\cite{mukherjee2018quantum,rose2018spectral}.

In order to efficiently simulate the stochastic process induced by the reset (see~Eq.~\eqref{eq:qwalk_reset}),
we use a Crank-Nicholson scheme~\cite{askar1978explicit}:
\begin{equation}
\ket{\phi^{n+1}}=
\begin{cases}
\ket{\phi(0)}  &\text{if}\, z\leq r\,\mathrm{d}t\,, \\
\ket{\phi^{n-1}}-2 \II \Delta t H_\mathrm{q} \ket{\phi^n}
  &\text{otherwise},
\end{cases}
\label{eq:crank_nicholson}
\end{equation}
where the superscript $n$ indicates the time step. For reset-CRWs, we use Eq.~\eqref{eq:qwalk_reset} and replace $\ket{\phi}$ by the probability vector $\bf p$ and $\II H_{\rm q}$ by the classical Hamiltonian $H_{\rm c}$. To numerically solve the evolution of CRWs, we use an Euler forward integration scheme.
\subsection{Classical-to-quantum walks}
\noindent
In Sec.~\ref{sec:qsws}, we introduced the quantum jump operators
\begin{equation}
 L_{nm}  = \sqrt{\epsilon}\sqrt{\bra{n}H_{\rm c} \ket{m} } \ket{n}\bra{m}
\end{equation}
to include the CRW in a quantum process. The resulting master equation 
of the QCW is Eq.~(\ref{eq:qsw}) and
the corresponding unravelling (see Eq.~(\ref{eq:sse}))
can be written as
\begin{align}
\begin{split}
  \ket{\D \phi}  
=  &-\II(1-\epsilon) H_{\rm q}  \ket{\phi} \D t\\
&+ \sum_{n,m\in V} \left[\frac{\sandwich{m}{\phi}}{
 \left|\sandwich{m}{\phi}\right|} e^{\II \frac{\pi}{2} \delta_{nm}} \ket{n} -\ket{\phi}\right] \D N_{nm}
 \end{split}
 \label{eq:c2q}
\end{align}
with the Kronecker delta $\delta_{nm}$. The quantum jump probability belonging to this stochastic 
Schr\"odinger equation reads
\begin{equation}
 P_{\rm j} = 
 \epsilon \D t  \bigg(1-\sum_{n\neq m \in V}
 \bra{n} H_{\rm c}\ket{m} \left| \sandwich{m}{\phi}\right|^2
     \bigg) = 2 \epsilon \D t
     \label{eq:pjcqw}
\end{equation}
where the first equality is the definition of the jump probability for the previously defined set of quantum jump
operators and the second equality is a direct consequence of the specific shape of the classical Hamiltonian
and the symmetry of the adjacency matrix. 
Note that the dissipative processes in Eq.~(\ref{eq:c2q}) do not contain
the interpolation parameter $\epsilon$. Rather the jump probability in Eq.~(\ref{eq:pjcqw}) is proportional to 
$\epsilon$, rendering quantum jumps impossible in the unitary limit $\epsilon\to0$.
\section{Numerical Results on graphs}
\label{sec:numerics}
\subsection{Classical and quantum walks with resetting}
\noindent
We now compare our analytical results of the stationary states of reset QCWs 
(see Eq.~\eqref{eq:statsolq} and \eqref{eq:statsolqQW})  with the 
corresponding numerical solutions, given an underlying reset process with 
rate $r$.
In Fig.~\ref{fig:comparison1}, we show the 
probabilities ${p_k}^\prime(r)$ and ${q_k}^\prime(r)$ that a 
node of degree $k$ is occupied by a classical and quantum random walker, 
respectively (see Eqs.~\eqref{eq:pk} and \eqref{eq:qk}). 
We perform simulations on Erd\H{o}s-R\'enyi, 
Barab\'asi-Albert, and peer-to-peer networks 
(see Sec.~\ref{sec:networks}). The degree 
distribution of Erd\H{o}s-R\'enyi networks is binomial 
whereas Barab\'asi-Albert networks and the peer-to-peer networks exhibit broader degree distributions (see Fig.~\ref{fig:networks}). As initial condition and reset state, we use a uniform distribution over all nodes. Thus, in the limit $r \rightarrow \infty$, the occupation probabilities satisfy ${p_k}^\prime(r)={q_k}^\prime(r)=\text{const.}$ for all degrees $k$. In other words, for the chosen reset protocol, classical and quantum walks share the same node occupation probabilities as $r\rightarrow\infty$. Our results show that solutions of Eqs.~\eqref{eq:statsolqQW} and \eqref{eq:pr_stat} agree well with the 
numerically-obtained occupation probabilities ${\bf p}^\prime(r)$ on all networks for 
different reset rates $r$. 

We observe that the underlying network structure has a significant effect on how the occupation probabilities of the CRW and QW approach the limiting uniform distribution as $r$ becomes larger. In the Supplemental Material~\cite{vimeo_walk}, we include an animation of the occupation-probability evolution for the three aforementioned networks. Quantum walks sample from the occupation probability distribution in a different way than CRWs. For an Erd\H{o}s-R\'enyi network and a reset rate $r=0.5$, low and high degree nodes are more likely to be sampled by a QW than by a CRW, which is different from what we observe for Barab\'asi-Albert and peer-to-peer networks (see Fig.~\ref{fig:comparison1}). For a given network, this difference in sampling can be controlled with the reset rate $r$. 
\begin{figure*}
\centering
\includegraphics[width=0.49\textwidth]{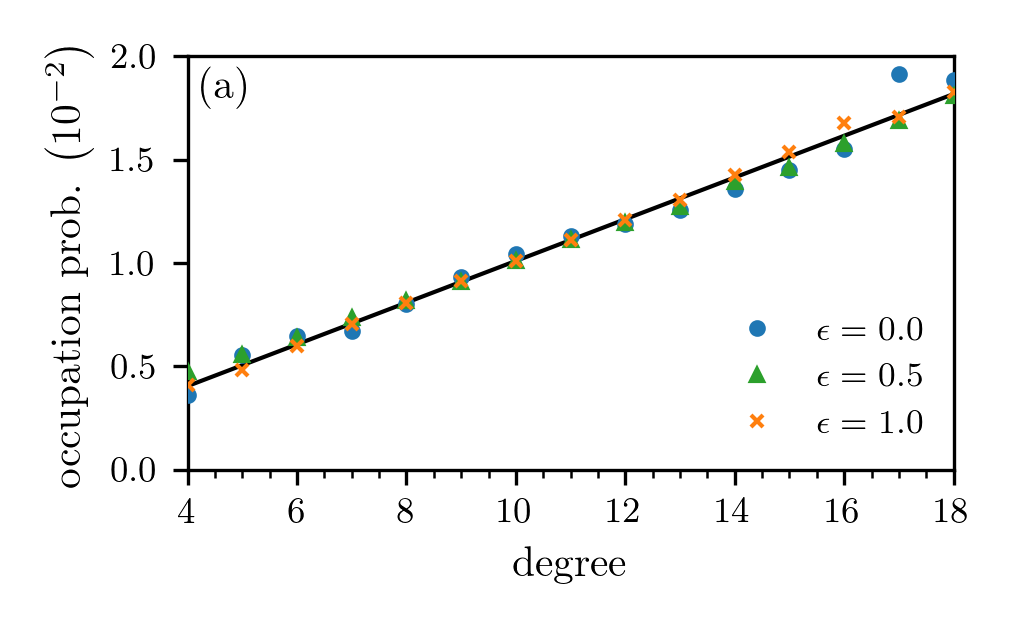}
\includegraphics[width=0.49\textwidth]{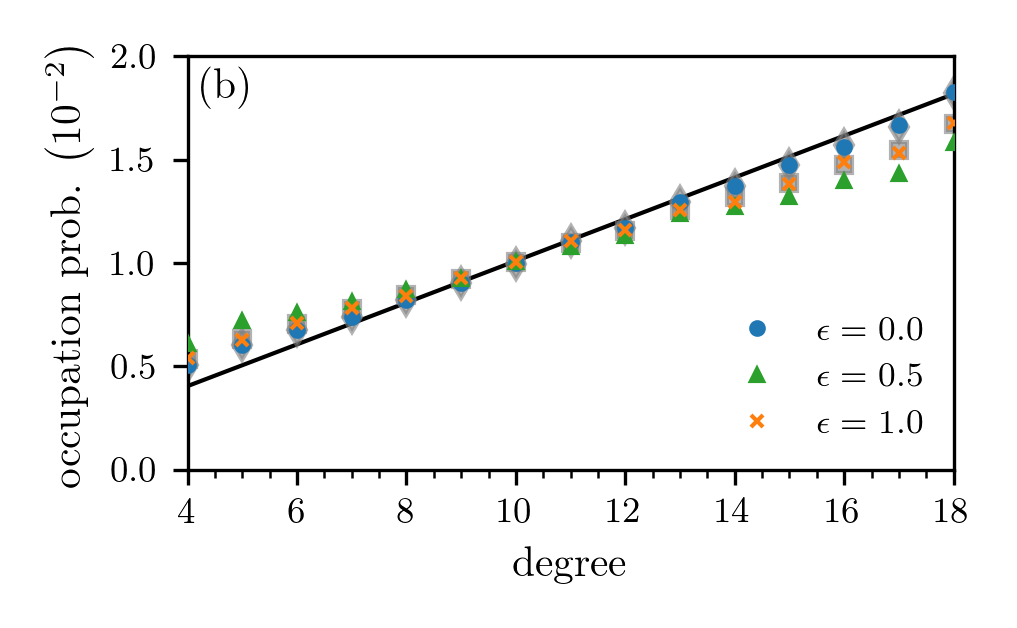}
\caption{\textbf{Occupation probability for quantum-to-classical walks.} 
We show the probability that a node of degree 
$k$ is occupied by a classical-to-quantum random walker for (a) $r=0$ and (b) $r=0.3$. As 
initial conditions and reset states, we use a uniform distribution over all nodes. Our numerical results are based on solutions of 
Eq.~\eqref{eq:c2q}. Simulations were performed on an Erd\H{o}s-R\'enyi network with $N=100$ nodes. The shown data points are averages over $2\times 10^5$ 
samples. The black solid line is the occupation probability of a CRW for $r=0$ and grey markers in panel (b) correspond to analytical solutions of a CRW ($\epsilon=1$) and QW $(\epsilon=0$).}
\label{fig:q2c}
\end{figure*}

To determine the difference between ${\bf p}^\prime(r)$ and ${\bf q}^\prime(r)$ as a function of $r$, we define the distance metric
\begin{equation}\label{eq:d}
d(r) =  \norm{{\bf p}^\prime(r)-{\bf q}^\prime(r)},
\end{equation}
where $\norm{\cdot}$ denotes the Euclidean norm. In Fig.~\ref{fig:comparison2}, we show $d(r)$ for the three networks of Sec.~\ref{sec:networks}. Numerical and analytical results are indicated by red and black solid lines, respectively. The distance between ${\bf p}^\prime(r)$ and ${\bf q}^\prime(r)$ is small, yet finite, for $r=0$ and vanishes as $r\rightarrow \infty$. Interestingly, we observe multiple inflection points in $d(r)$ for the Erd\H{o}s-R\'enyi and Barab\'asi-Albert networks and a maximum distance $d(r)$ at $r \approx 0.3$ for the peer-to-peer
and Erd\H{o}s-R\'enyi networks. After the initial local minimum at $r \approx 0$, the distance between the classical and quantum occupation-probability distributions reaches a second pronounced local minimum at $r\approx 1$ and $r \approx 1.5$ for the Erd\H{o}s-R\'enyi and Barab\'asi-Albert networks, respectively. Each minimum corresponds to a cross-over of ${\bf p}^\prime(r)$ and ${\bf q}^\prime(r)$.
Based on the observed behavior of ${\bf p}^\prime(r)$, ${\bf q}^\prime(r)$, and $d(r)$, we conclude that classical and quantum occupation probabilities strongly depend on the underlying network structure and are not affected in the same way by changes of the reset rate $r$. 
The initial increase of the distance $d(r)$ with the reset rate seems to be universal throughout our results and may be explained with the initially strong impact of stochastic resetting on CRWs. While the long-time behavior of QWs is characterized by Eq.~(\ref{eq:cesaro}), CRWs have a unique stationary distribution that differs from our chosen reset state in heterogeneous networks. Therefore,
the resetting mechanism introduces a strong competition between the reset state and the $r=0$ stationary state that strongly affects
the node occupation properties of CRWs.
Conversly, for large reset rates, the results in Fig.~\ref{fig:comparison1} and \cite{vimeo_walk} show that QWs are closer to the uniform reset state for large reset rates.
This can be qualitatively traced back to the strong mixing of the unitary dynamics: As the reset becomes the dominant contribution in the dynamics, mixing leads to 
a rather large overlap with the reset state and thus facilitates the relaxation into the reset state.

As described in Ref.~\cite{faccin2013degree}, differences between the node occupation statistics of CRWs and QWs (without stochastic resetting, i.e., $r=0$) are largest on networks with a heterogeneous degree distribution (e.g., Barab\'asi-Albert networks). As we show in Fig.~\ref{fig:comparison1}, stochastic resetting can substantially increase these differences. For instance, in the considered Barab\'asi-Albert network, the distance $d(r)$ between CRWs and QWs is about 4--5 times larger for certain reset rates than $d(r=0)$.

\subsection{Reset quantum-to-classical walks}
\noindent
Figures~\ref{fig:q2c} (a) and (b) show the occupation probability distribution of a QCW (see Eq.~\eqref{eq:c2q}) with $r=0,0.3$ and $\epsilon=0,0.5,1$ on an Erd\H{o}s-R\'enyi network. Note that the limits $\epsilon = 0$ and $\epsilon = 1$ correspond to the purely quantum and classical case, respectively. For $\epsilon=0.5$, we obtain a hybrid quantum-to-classical walk with yet different node-occupation statistics. 

Our simulation results in Fig.~\ref{fig:q2c} (a) confirm that the node occupation statistics of a QCW with $\epsilon=1$ and $r=0$ (orange crosses) agree with those of a CRW (black solid line). We also find that the analytical solutions \eqref{eq:statsolqQW} and \eqref{eq:pr_stat} of the limiting cases $\epsilon = 0,1$ (grey markers) agree with the corresponding simulations results for a reset rate of $r=0.3$ (see Fig.~\ref{fig:q2c}(b)). Interestingly, the hybrid quantum-classical walk with $\epsilon=0.5$ is affected more by the finite reset rate $r=0.3$ than its purely classical and quantum counterparts.
\section{Discussion and Outlook}
\label{sec:discussion}
\noindent
Random walks are important models of diffusive processes in many branches of science. In this work, we introduced a framework for the study of classical, quantum, and hybrid random walks with stochastic resetting on networks. We derived analytical solutions for the probability that a classical or quantum random walker occupies a certain node on a network. These analytical results, valid for general reset rates and network structures, are in perfect accordance with numerical solutions of the underlying master equations. Our results also revealed differences in the way classical and quantum walks with reset sample nodes with certain degrees. Both walks react differently to changes in the reset rate, which may be used as a control parameter to achieve desired node occupation (or sampling) statistics in quantum search and optimization algorithms~\cite{childs2004spatial,douglas2008classical,chakraborty2016spatial}. 

\textcolor{black}{For the networks and uniform reset state that we studied in this paper, quantum walks are closer to the reset state than classical walks as the reset rate $r$ becomes large (see Fig.~\ref{fig:comparison1} and \cite{vimeo_walk}). This behavior is linked to the unitary dynamics of QWs, which mixes quantum states (see Eq.~\eqref{eq:statsolqQW}) and produces wave-function components that are close to the uniform reset state. Therefore, QWs reach such states ``faster'' (in the terms of a smaller reset rate) than classical walks. For large reset rates, the stationary distribution of CRWs is significantly altered from the one without resets. We thus find a competition between random-walk dynamics and reset dynamics, where both processes \textcolor{black}{compete} at different time scales.}

In future work, our framework may be used to study how quantum walk search is affected by stochastic resetting as was done for classical walks in Ref.~\cite{riascos2020random}.
\textcolor{black}{Our framework is also directly applicable to compute classical, quantum, and hybrid random-walk-based centrality metrics on networks. In contrast to earlier studies~\cite{sanchez2012quantum,rossi2014node,izaac2017centrality}, we also account for stochastic resetting, providing a possibility to reach desired node occupation statistics and design tailored centrality measures.}

\textcolor{black}{
Other areas of application for resetting mechanisms include quantum feedback 
control~\cite{Nel00,East17} and repeated measurements~\cite{evans2020stochastic}.
In these setups one repeatedly 
measures the quantum system and tailors the dynamics depending on the experimental
outcome. Such formulations might be useful to impose constraints on the quantum system
and in this manner induce non-trivial dynamical behavior on otherwise free quantum
systems~\cite{Mara15,Wald16,Wald18,Timpa19,Wald20,Hali20}.
}

\textcolor{black}{While there has been rapid theoretical progress in the field of classical and quantum dynamics with stochastic resetting~\cite{evans2020stochastic}, it will be important to also focus on experimental realizations of such processes in future studies and adapt theoretical models according to experimental protocols. As described in Ref.~\cite{evans2020stochastic}, ``in a theoretical model one often assumes instantaneous resetting which is impossible to achieve experimentally. Thus experimentalists need to devise different types of resetting protocols, which in turn pose interesting theoretical challenges.'' }

%
All codes are publicly available on GitHub~\cite{git_walk}.
\section*{Acknowledgments}
We thank M Henkel, M Heyl, and F Semi\~ao for helpful comments. LB acknowledges financial support from the SNF Early Postdoc.Mobility fellowship on ``Multispecies interacting
stochastic systems in biology'' and from the Army Research Office (W911NF-18-1-0345).
\vfill
%
%
%
%
%
%
\appendix
\renewcommand\thefigure{\thesection.\arabic{figure}}
\setcounter{figure}{0}    
\section{Inclusion of CRW in QCW}
\label{app:eps1}
\noindent
In this appendix we show that the construction outlined in Sec.~\ref{sec:qsws2} reduces to the 
CRW (see Eq.~(\ref{eq:crw})) in the limit $\epsilon\to1$.
This is done by evaluating the time evolution of the corresponding 
occupation probabilities $p_\ell(t)= \bra{\ell} \varrho \ket{\ell}$.
We utilize the Lindblad master equation~(\ref{eq:qsw}) to derive the time evolution of the
probabilities, viz.
\begin{align}
    \begin{split}
 \frac{\D p_\ell}{\D t}
&=\sum_{n\neq m} \bra{n}H_{\rm c}\ket{m} \big[
\frac{\varrho_{m\ell}+\varrho_{\ell m}}{2}  \delta_{m\ell}
-p_m \delta_{n\ell} 
\big]
\end{split}
\end{align}
It is then clear that we can reduce the double sum to a single 
sum due to the Kronecker delta terms and we see that there are 
two distinct contributions, viz.
\begin{align}
\begin{split}
 \frac{\D p_\ell}{\D t}&=\sum_{m\neq \ell}-\bra{\ell}H_{\rm c}\ket{m} p_m
+\sum_{n\neq \ell}\bra{n}H_{\rm c}\ket{\ell} 
    p_{\ell}
    \end{split}
\end{align}
The first term is already in the correct shape to reproduce the classical dynamics and solely
the diagonal contributions are differ. We need to use the explicit form of the classical Hamiltonian, in 
particular that the diagonal elements are equal to $1$ in order to rewrite the second contribution.
It is then straightforward to carry the calculation out as
\begin{align}
\begin{split}
 \frac{\D p_\ell}{\D t}&=\sum_{m\neq \ell}-\bra{\ell}H_{\rm c}\ket{m} p_m
-\sum_{n\neq \ell}\frac{A_{n\ell}}{k_\ell}
    p_{\ell}\\
&=\sum_{m\neq \ell}-\bra{\ell}H_{\rm c}\ket{m} p_m
-p_{\ell}\\
&
= 
\sum_{m}-\bra{\ell}H_{\rm c}\ket{m} p_m
    \end{split}
\end{align}
where we also used the symmetry of the adjacency matrix $A_{nm}= A_{mn}$.
This proves that the generated dynamics coincides with the stochastic
dynamics defined in Eq.~(\ref{eq:crw}).
\section{Resolvent of infinitesimal generator}
\label{app:laplace}
\noindent
In this appendix, we illustrate the use of resolvents to formally solve
differential equations. We consider the matrix differential equation
\begin{equation}
\dot{\mathbf{x}}=A \mathbf{x}\,.
\label{eq:matrix_differential_eq}
\end{equation}
for the vector function $\mathbf{x} = (x_1,\dots,x_d)$ and $A \in \mathbb{R}^{d\times d}$.
We introduce the Laplace transform $\hat{\mathbf{x}}(s)$ of $\mathbf{x}(t)$ as
\begin{equation}
\hat{\mathbf{x}}(s)=\int_0^\infty \mathbf{x}(t) e^{-s t} \, \mathrm{d}t\,.
\end{equation}
Applying the Laplace transform to Eq.~\eqref{eq:matrix_differential_eq} yields
\begin{equation}
\hat{\mathbf{x}}(s)s-\mathbf{x}(0)=A \hat{\mathbf{x}}(s)\, .
\label{eq:Xs}
\end{equation}
In this way we reduced the differential equation~(\ref{eq:matrix_differential_eq}) to an
algebraic equation
\begin{equation}
 \hat{\mathbf{x}}(s)=(\mathbb{1} s-A)^{-1} \mathbf{x}(0) \,,
\end{equation}
whose solution is readily found.
The operator $(\mathbb{1} s-A)^{-1}$ is called the \emph{resolvent} of 
$A$~\cite{pazy2012semigroups}. 
An integral representation of the resolvent is found by
applying the Laplace transform directly to the 
solution of Eq.~\eqref{eq:matrix_differential_eq}, we find
\begin{equation}
\hat{\mathbf{x}}(s) = \int_0^\infty e^{-st} e^{A t} \mathbf{x}(0) \,\mathrm{d}t\,.
\label{eq:laplace_solution}
\end{equation}
Combining Eqs.~\eqref{eq:Xs} and \eqref{eq:laplace_solution} yields
\begin{equation}
\int_0^\infty e^{-st} e^{A t} \mathbf{x}(0) \,\mathrm{d}t=(\mathbb{1} s-A)^{-1} 
\mathbf{x}(0)\,.
\end{equation}
This identity is used in Eq.~\eqref{eq:pr_stat} in order 
to explicitly determine the stationary state of the CRW.
%
%
%
\bibliography{refs}
\bibliographystyle{apsrev4-1}
\end{document}